# Интервальные реберные раскраски "лестниц Мебиуса"

П.А. Петросян

Институт проблем информатики и автоматизации НАН РА,
г. Ереван, ул. П. Севака 1, 375014, Армения, e-mail: pet_petros@yahoo.com

Показано, что "лестницы Мебиуса" интервально окрашиваемы. Найдены все значения $t$, для которых существует интервальная реберная $t$-раскраска данных графов.

Пусть $G = (V, E)$ - неориентированный связный граф без кратных ребер и петель [1], $V(G)$ - множество вершин графа $G$, $E(G)$ - множество ребер графа $G$. Степень вершины $x$ в $G$ обозначим через $d_G(x)$, максимальную из степеней вершин через $\Delta(G)$, хроматический класс [2] $G$ через $\chi'(G)$. Граф назовем регулярным, если все его вершины имеют одну и ту же степень. Расстояние между вершинами $x$ и $y$ обозначим через $d(x,y)$, а диаметр графа $G$ - через $d(G)$. Если $\alpha$ - правильная раскраска [3] ребер графа $G$, то цвет ребра $e \in E(G)$ при раскраске $\alpha$ обозначим через $\alpha(e)$.

Интервальной [4] реберной $t$-раскраской графа $G$ назовем правильную раскраску ребер $G$ в цвета $1, 2, \ldots, t$, при которой в каждый цвет $i, 1 \leq i \leq t$, окрашено хотя бы одно ребро $e_i \in E(G)$, и ребра, инцидентные каждой вершине $x \in V(G)$, окрашены в $d_G(x)$ последовательных цветов.

Граф $G$ назовем интервально окрашиваемым, если существует $t \geq 1$, для которого $G$ имеет интервальную реберную $t$-раскраску.

Множество всех интервально окрашиваемых графов обозначим через $\mathfrak{N}$ [5].

Для $G \in \mathfrak{N}$ обозначим через $w(G)$ и $W(G)$, соответственно, наименьшее и наибольшее $t$, для которого $G$ имеет интервальную реберную $t$-раскраску.

В [6] доказано, что задача распознавания: $G \in \mathfrak{N}$ или $G \notin \mathfrak{N}$, является $NP$-полной [7,8] для двудольных графов $G$

Теорема 1 [9]. Если $G$ - двудольный граф и $G \in \mathfrak{N}$, то
$$W(G) \leq d(G)(\Delta(G) - 1) + 1.$$

Теорема 2 [9]. Если $G$ содержит цикл нечетной длины и $G \in \mathfrak{N}$, то
$$W(G) \leq (d(G) + 1)(\Delta(G) - 1) + 1.$$

В [5] доказана

Теорема 3. Пусть $G$ - регулярный граф.
(1) $G \in \mathfrak{N}$ тогда и только тогда, когда $\chi'(G) = \Delta(G)$.
(2) Если $G \in \mathfrak{N}$ и $\Delta(G) \leq t \leq W(G)$, то $G$ имеет интервальную реберную $t$-раскраску.

Известно [2], что для любого графа $G$  $\Delta(G) \leq \chi'(G) \leq \Delta(G) + 1$. В [10] доказано, что для регулярного графа $G$ проблема определения: $\chi'(G) = \Delta(G)$ или $\chi'(G) \neq \Delta(G)$, является $NP$-полной. Отсюда и из теоремы 3 следует, что для регулярного графа $G$ проблема определения: $G \in \mathfrak{N}$ или $G \notin \mathfrak{N}$, является также $NP$-полной.

Настоящая работа посвящена исследованию интервальных реберных раскрасок "лестниц Мебиуса". Не определяемые понятия можно найти в [1,3,5,11].

**Лемма 1.** При $n \geq 2$  $d(M_{2n}) = \lceil \frac{n}{2} \rceil$.

**Доказательство.** Пусть $V(M_{2n}) = \{x_1, x_2, \ldots, x_{2n}\}$,

$E(M_{2n}) = \{(x_i, x_{i+1}) | 1 \leq i \leq 2n-1\} \cup \{(x_{2n}, x_1)\} \cup \{(x_i, x_{n+i}) | 1 \leq i \leq n\}$.

Ясно, что $d(x_1, x_{n + \lceil \frac{n}{2} \rceil}) = \lceil \frac{n}{2} \rceil$. Отсюда следует, что $d(M_{2n}) \geq \lceil \frac{n}{2} \rceil$.

Докажем, что $d(M_{2n}) \leq \lceil \frac{n}{2} \rceil$.

Очевидно, для этого достаточно показать, что при $1 \leq i < j \leq 2n$ $d(x_i, x_j) \leq \lceil \frac{n}{2} \rceil$.

**Случай 1.** $1 \leq i < j \leq n$.

Покажем, что $d(x_i, x_j) \leq \lceil \frac{n}{2} \rceil$. Если $j - i \leq \lceil \frac{n}{2} \rceil$, то $d(x_i, x_j) \leq \lceil \frac{n}{2} \rceil$.

Пусть теперь $j - i \geq \lceil \frac{n}{2} \rceil + 1$. Рассмотрим простую цепь

$P_1 = (x_i, (x_i, x_{i-1}), x_{i-1}, \ldots, x_2, (x_2, x_1), x_1, (x_1, x_{2n}), x_{2n}, (x_{2n}, x_n), x_n, (x_n, x_{n-1}), x_{n-1}, \ldots, x_{j+1}, (x_{j+1}, x_j), x_j)$.

Легко видеть, что $d(x_i, x_j) \leq i - 1 + 2 + n - j = n + 1 - (j - i) \leq n - \lceil \frac{n}{2} \rceil = \lfloor \frac{n}{2} \rfloor$.

**Случай 2.** $n + 1 \leq i < j \leq 2n$.

Покажем, что $d(x_i, x_j) \leq \lceil \frac{n}{2} \rceil$. Если $j - i \leq \lceil \frac{n}{2} \rceil$, то $d(x_i, x_j) \leq \lceil \frac{n}{2} \rceil$.

Пусть теперь $j - i \geq \lceil \frac{n}{2} \rceil + 1$. Рассмотрим простую цепь

$P_2 = (x_i, (x_i, x_{i-1}), x_{i-1}, \ldots, x_{n+2}, (x_{n+2}, x_{n+1}), x_{n+1}, (x_{n+1}, x_n), x_n, (x_n, x_{2n}), x_{2n}, (x_{2n}, x_{2n-1}), x_{2n-1}, \ldots, x_{j+1}, (x_{j+1}, x_j), x_j)$.

Легко видеть, что $d(x_i, x_j) \leq i - n - 1 + 2 + 2n - j = n + 1 - (j - i) \leq n - \lceil \frac{n}{2} \rceil = \lfloor \frac{n}{2} \rfloor$.

**Случай 3.** $1 \leq i \leq n$, $n + 1 \leq j \leq 2n$, $n + i \leq j$.

Покажем, что $d(x_i, x_j) \leq \lceil \frac{n}{2} \rceil$. Если $j - i - n + 1 \leq \lceil \frac{n}{2} \rceil$, то $d(x_i, x_j) \leq \lceil \frac{n}{2} \rceil$.

Пусть теперь $j - i - n \geq \lceil \frac{n}{2} \rceil$. Рассмотрим простую цепь

$P_3 = (x_i, (x_i, x_{i-1}), x_{i-1}, \ldots, x_2, (x_2, x_1), x_1, (x_1, x_{2n}), x_{2n}, (x_{2n}, x_{2n-1}), x_{2n-1}, \ldots, x_{j+1}, (x_{j+1}, x_j), x_j)$

Легко видеть, что $d(x_i, x_j) \leq i - 1 + 1 + 2n - j = 2n - (j - i) \leq n - \lceil \frac{n}{2} \rceil = \lfloor \frac{n}{2} \rfloor$.

**Случай 4.** $1 \leq i \leq n$, $n + 1 \leq j \leq 2n$, $n + i \geq j$.

Покажем, что $d(x_i, x_j) \leq \lceil \frac{n}{2} \rceil$. Если $n + i - j + 1 \leq \lceil \frac{n}{2} \rceil$, то $d(x_i, x_j) \leq \lceil \frac{n}{2} \rceil$.

Пусть теперь $n + i - j \geq \lceil \frac{n}{2} \rceil$. Рассмотрим простую цепь

$P_4 = (x_i, (x_i, x_{i+1}), x_{i+1}, \ldots, x_{n-1}, (x_{n-1}, x_n), x_n, (x_n, x_{n+1}), x_{n+1}, \ldots, x_{j-1}, (x_{j-1}, x_j), x_j)$.

Легко видеть, что $d(x_i, x_j) \leq n - i + 1 + j - n - 1 = j - i \leq n - \lceil \frac{n}{2} \rceil = \lfloor \frac{n}{2} \rfloor$.
Лемма 1 доказана.

Лемма 2. При $n \geq 2$  $W(M_{2n}) \geq n + 2$.
Доказательство. Пусть $V(M_{2n}) = \{x_1, x_2, \ldots, x_{2n}\}$,

$E(M_{2n}) = \{(x_i, x_{i+1}) | 1 \leq i \leq 2n - 1\} \cup \{(x_{2n}, x_1)\} \cup \{(x_i, x_{n+i}) | 1 \leq i \leq n\}$.

Определим раскраску $\alpha$ рёбер графа $M_{2n}$.

Случай 1. $n = 2m, m \in N$.
   Для $i = 1, 2, \ldots, m$ положим

$$\alpha((x_{m-1+i}, x_{3m-1+i})) = 2i - 1.$$

Для $i = 1, 2, \ldots, m - 1$ положим

$$\alpha((x_{m-i}, x_{3m-i})) = 2(i + 1).$$

Для $i = 1, 2, \ldots, m$ положим

$$\alpha((x_{m-1+i}, x_{m+i})) = \alpha((x_{3m-1+i}, x_{3m+i})) = 2i.$$

Для $i = 1, 2, \ldots, m - 1$ положим

$$\alpha((x_{m-i}, x_{m+1-i})) = \alpha((x_{3m-i}, x_{3m+1-i})) = 2i + 1 \text{ и}$$
$$\alpha((x_1, x_{4m})) = \alpha((x_{2m}, x_{2m+1})) = 2m + 1, \; \alpha((x_{2m}, x_{4m})) = 2m + 2.$$

Случай 2. $n = 2m + 1, m \in N$.
   Для $i = 1, 2, \ldots, m + 1$ положим

$$\alpha((x_{m+i}, x_{3m+1+i})) = 2i - 1.$$

Для $i = 1, 2, \ldots, m - 1$ положим

$$\alpha((x_{m+1-i}, x_{3m+2-i})) = 2(i + 1).$$

Для $i = 1, 2, \ldots, m$ положим

$$\alpha((x_{m+i}, x_{m+1+i})) = \alpha((x_{3m+1+i}, x_{3m+2+i})) = 2i.$$

Для $i = 1, 2, \ldots, m$ положим

$$\alpha((x_{m+1-i}, x_{m+2-i})) = \alpha((x_{3m+2-i}, x_{3(m+1)-i})) = 2i + 1 \text{ и}$$
$$\alpha((x_1, x_{4m+2})) = \alpha((x_{2m+1}, x_{2m+2})) = 2m + 2, \; \alpha((x_1, x_{2m+2})) = 2m + 3.$$

Легко видеть, что $\alpha$ является интервальной реберной $(n + 2)$-раскраской графа $M_{2n}$ при $n \geq 2$.

Лемма 2 доказана.

Теорема 4. При $n \geq 2$
(1) $M_{2n} \in \mathfrak{N}$,
(2) $w(M_{2n}) = 3$,
(3) $W(M_{2n}) = n + 2$,
(4) если $w(M_{2n}) \leq t \leq W(M_{2n})$, то $M_{2n}$ имеет интервальную реберную $t$-раскраску.

Доказательство. Так как $M_{2n}$ при $n \geq 2$ является регулярным графом, удовлетворяющим равенству $\chi'(M_{2n}) = \Delta(M_{2n}) = 3$, то из теоремы 3 вытекают утверждения (1) и (2) доказываемой теоремы.

Докажем, что при $n \geq 2$ $W(M_{2n}) = n + 2$.

Из леммы 2 следует, что при $n \geq 2$ $W(M_{2n}) \geq n + 2$.

Покажем, что при $n \geq 2$ $W(M_{2n}) \leq n + 2$.

Случай 1. $n = 2m + 1$, $m \in N$.

Так как граф $M_{4m+2}$ является двудольным, то из теоремы 1 и леммы 1 вытекает неравенство $W(M_{4m+2}) \leq 2m + 3$.

Случай 2. $n = 2m$, $m \in N$.

Так как граф $M_{4m}$ содержит цикл нечетной длины, то из теоремы 2 и леммы 1 вытекает неравенство $W(M_{4m}) \leq 2m + 3$.

Покажем, что $W(M_{4m}) \leq 2m + 2$.

Пусть $V(M_{4m}) = \{x_1, x_2, \ldots, x_{4m}\}$,
$E(M_{4m}) = \{(x_i, x_{i+1})| 1 \leq i \leq 4m - 1\} \cup \{(x_{4m}, x_1)\} \cup \{(x_i, x_{2m+i})| 1 \leq i \leq 2m\}$.

Предположим, что $\alpha$ является интервальной реберной $(2m + 3)$-раскраской графа $M_{4m}$.

Пусть $e \in E(M_{4m})$ и $\alpha(e) = 2m + 3$.

Случай 2a). $e \in \{(x_i, x_{i+1})| 1 \leq i \leq 4m - 1\} \cup \{(x_{4m}, x_1)\}$.

Без ограничения общности можно считать, что $e = (x_1, x_2)$ и существует вершина $x_k$, где $2 \leq k \leq 2m + 1$, которая инцидентна ребру цвета 1. Рассмотрим простую цепь

$$P_1 = (x_2, (x_2, x_3), x_3, \ldots, x_{k-1}, (x_{k-1}, x_k), x_k).$$

Из интервальности раскраски $\alpha$ следует, что $k \geq m + 2$. Теперь рассмотрим простую цепь

$$P_2 = (x_k, (x_k, x_{k+1}), x_{k+1}, \ldots, x_{2m}, (x_{2m}, x_{2m+1}), x_{2m+1}).$$

Из интервальности раскраски $\alpha$ следует, что $\alpha((x_{2m}, x_{2m+1})) \leq 2m - 2$. Так как $\alpha((x_1, x_{2m+1})) \geq 2m + 1$, то $\alpha((x_{2m}, x_{2m+1})) \geq 2m - 1$, что приводит к противоречию, и доказательство в случае 2a) завершается.

Случай 2б). $e \in \{(x_i, x_{2m+i})| 1 \leq i \leq 2m\}$.

Без ограничения общности можно считать, что $e = (x_1, x_{2m+1})$ и существует вершина $x_k$, где $2 \leq k \leq 2m + 1$, которая инцидентна ребру цвета 1. Рассмотрим простую цепь

$$P_3 = (x_1, (x_1, x_2), x_2, \ldots, x_{k-1}, (x_{k-1}, x_k), x_k).$$

Из интервальности раскраски $\alpha$ следует, что $k \geq m + 1$. Теперь рассмотрим простую

цепь

$$P_4 = (x_k, (x_k, x_{k+1}), x_{k+1}, \ldots, x_{2m}, (x_{2m}, x_{2m+1}), x_{2m+1}).$$

Из интервальности раскраски $\alpha$ следует, что $\alpha((x_{2m}, x_{2m+1})) \leq 2m$. Так как $\alpha((x_1, x_{2m+1})) = 2m + 3$, то $\alpha((x_{2m}, x_{2m+1})) \geq 2m + 1$, что приводит к противоречию, и доказательство в случае 2б) завершается.

Для завершения доказательства теоремы 4 достаточно показать, что если $3 \leq t \leq n + 2$, то $M_{2n}$ имеет интервальную реберную $t$-раскраску, а это непосредственно следует из теоремы 3.

Теорема 4 доказана.